\documentclass[aps,prl,reprint,groupedaddress]{revtex4-1}
\usepackage{graphicx} 
\usepackage{dcolumn} 
\usepackage{bm}
\usepackage{hyperref}

\begin{document}

\title{Hyperon Dynamics in Heavy-ion Collisions near threshold energy}
\author{Ding-Chang Zhang, Hui-Gan Cheng, Zhao-Qing Feng }
\email{fengzhq@scut.edu.cn}
\affiliation{School of Physics and Optoelectronic Technology, South China University of Technology, Guangzhou 510641, China}

\date{\today}

\begin{abstract}
Within the framework of quantum molecular dynamics transport model, the isospin and in-medium effects on the hyperon production in the reaction of $^{197}$Au + $^{197}$Au are investigated thoroughly. A repulsive hyperon-nucleon potential from the chiral effective field theory is implemented into the model, which is related to the hyperon momentum and baryon density. The correction on threshold energy of the elementary hyperon cross section is taken into account. It is found that the $\Sigma$ yields are suppressed in the domain of midrapidity and kinetic energy spectra with the potential. The hyperons are emitted in the reaction plane because of the strangeness exchange reaction and reabsorption process in nuclear medium. The $\Sigma^{-}/\Sigma^{+}$ ratio depends on the stiffness of nuclear symmetry energy, in particular in the high-energy region (above 500 MeV).

\begin{description}
\item[PACS number(s)]
21.65.Ef, 25.75.Dw, 25.75.Ld
\end{description}
\end{abstract}

\maketitle

Heavy-ion collisions provide good environment for studying the dense hadronic matter and in-medium properties of hadron, especially for strange particle production, which has attracted much attention. The hyperons might be created below the threshold energy in the heavy-ion collisions. The hyperon-nucleon (YN) interaction in theories \cite{Ha01,Da02,Fu03,Fu04} and YN scattering in experiments \cite{Na05,Ka06,Ko07} have been studied, but the interaction of strange particles in dense nuclear matter and the dynamics of hyperon production in high-density region are not well understood up to now. In pioneer works, the YN potential calculated by relativistic mean-field model (RMF) are generally attractive \cite{Ru08,Lo09,Ba10}, but a weak repulsive $\Sigma$N potential was argued \cite{Ga11,Do12}. Recently, the repulsive $\Sigma$N potential has been calculated by Nijmegen extended-soft-core (ESC) potential model \cite{Th13} and chiral effective field theory (EFT) \cite{Pe14}. In particular, some relatively precise parameters of $\Sigma$N potential that is strongly isospin-dependent are obtained by EFT \cite{Ha15}, which will influence the strange particle dynamics in heavy-ion collisions \cite{Ga16,lv17,He18}.

The equation of state in isospin asymmetric dense nuclear matter, namely the high-density symmetry energy, is not very clear up to now, which plays an important role in studying the properties and structures of neutron stars \cite{Xu19}. It is generally believed that hyperons will appear inside the neutron stars \cite{We20} in the density region of 2-3$\rho_{0}$. The appearance of strange particles leads to softening the nuclear equation of state and less mass of neutron star\cite{Ba21}. The ingredients of hyperons in neutron stars are associated with the YN potentials. For instance, the attractive $\Sigma$N potential results in the appearance of $\Sigma$ in the low-density region in neutron stars before $\Lambda$. However, the repulsive $\Sigma$N potential leads to the disappearance of $\Sigma$ \cite{Sc22}. The pressure gradient of the high-density nuclear matter generated via the heavy-ion collisions in high energy influences the hyperon emission in phase space. The collective flows of the ejected particles in heavy-ion collisions is related to the nuclear equation of state. It is a topical approach to extract the information of the high-density equation of state from the particle emission \cite{Ya23}.

The strangeness nuclear physics has been planned as one of topical issues by the large-scale scientific facilities in the world, for instance, High Intensity heavy-ion Accelerator Facility (HIAF) in China \cite{Ya24}, Alternating Gradient Synchrotron (AGS) at BNL in the United State \cite{Pi25}, the LHC-ALICE at European Organization for Nuclear Research \cite{Do26}, Japan Proton Accelerator Research Complex (J-PARC) \cite{Ta27} etc. In theory, several models have been developed to describe the strange particle dynamics in heavy-ion collisions, such as the statistical model \cite{An28}, intranuclear cascade model \cite{Cu29}, transport models \cite{Ha30,Bu31}. The intensive investigation on the hyperon dynamics in heavy-ion collisions is helpful for extracting the high-density symmetry energy and YN interaction (potential, scattering etc).

In this letter, the in-medium properties of $\Sigma$ hyperons and nuclear symmetry energy in the domain of high-baryon densities are to be investigated with the Lanzhou quantum molecular dynamics (LQMD) transport model. The production of resonances with the mass below 2 GeV, hyperons ($\Lambda$, $\Sigma$, $\Xi$) and mesons ($\pi$, $\eta$, $K$, $\overline{K}$, $\rho$, $\omega$) is coupled in the reaction channels via meson-baryon and baryon-baryon collisions \cite{Fe32,Fe33}. The temporal evolutions of nucleons are described by Hamilton's equations of motion under the self-consistently generated two-body and three-body interaction potential with the Skyrme interaction. At the considered energies, the hyperons are mainly contributed from the channels of baryon-baryon and pion-baryon collisions. The symmetry energy in the LQMD model is composed of the kinetic energy from fermionic motion, the local interaction and the momentum potential at the saturation density with the value of 31.5 MeV \cite{Fe34}. The stiffness of symmetry energy is adjusted by the local part $E_{sym}^{loc}=\frac{1}{2} C_{sym}(\rho/\rho_{0})^{\gamma_{s}}$ with $C_{sym}$ being 38 MeV. The parameter $\gamma_{s}$ is chosen to get the suitable case from constraining the isospin observables, e.g., the values of 0.5, 1 and 2 being the soft, linear and hard symmetry energy, respectively.

The evolution of hyperons is also determined by the Hamiltonian, which is given by
\begin{eqnarray}
H_{Y}=\sum^{N_{Y}}_{i=1}(V^{Coul}_{i}+V^{Y}_{opt}(\bm{p}_{i},\rho_{i})+\sqrt{\bm{p}^{2}_{i}+m^{2}}).
\end{eqnarray}
where the $N_{Y}$ is the total number of hyperons including charged resonances. Here the Y could be $\Lambda$, $\Sigma$ or $\Xi$ hyperons in our model. Here the Coulomb interaction $V^{Coul}_{i}$ of hyperon and charged baryons takes the form of point charge. The $\lambda$ potential is calculated on the basis of the light-quark counting rule. The self-energies of hyperons are assumed to be two thirds of that experienced by nucleons, which leads to the $\Lambda-$nucleon potential in nuclear matter at the saturation density being the value of -32 MeV \cite{Fe35}.
The $\Sigma$N optical potential is estimated by
\begin{eqnarray}
V^{\Sigma}_{opt}(\bm{p}_{i},\rho_{i}) = && V_{0}(\rho_{i}/\rho_{0})^{\gamma_{s}}+V_{1}(\rho_{n} - \rho_{p})t_{\Sigma}\rho_{i}^{\gamma_{s}-1}/\rho_{0}^{\gamma_{s}}        \nonumber\\
&& + C_{mom}\rho_{i}\ln(\epsilon\bm{p}^{2}_{i}+1).
\end{eqnarray}
Here, the isospin quantities are taken as $\bm{t}_{\Sigma}$=1, 0, and -1 for $\Sigma^{-}$, $\Sigma^{0}$ and $\Sigma^{+}$, respectively. The $C_{mom}$ and $\epsilon$ are 1.76 MeV and 500 $c^{2}/$GeV$^{2}$. The values of 14.8 MeV and 67.8 MeV for the isoscalar $V^{\Sigma}_{0}$ and isovector $V^{\Sigma}_{1}$ parts are taken by fitting the results from the next-to-leading order (NLO) in chiral EFT with a cutoff of 600 MeV \cite{Ha15}. The hyperon dynamics in heavy-ion collisions is influenced by the optical potential. Shown in the FIG. 1 is the $\Sigma$ optical potential as functions of density and momentum. The difference of the potentials with the isospin splitting is obvious with the baryon density. The effective mass $m^{\ast}_{Y}=V^{Y}_{opt}(p=0,\rho_{i})+m_{Y}$ is implemented into the threshold energy correction for the hyperon creation. It is obvious that the repulsive potential will enhance the threshold energy and consequently results in the reduction of hyperon production. On the contrary, the attractive potential manifests an opposite contribution.

\begin{figure}
\includegraphics[width=8 cm]{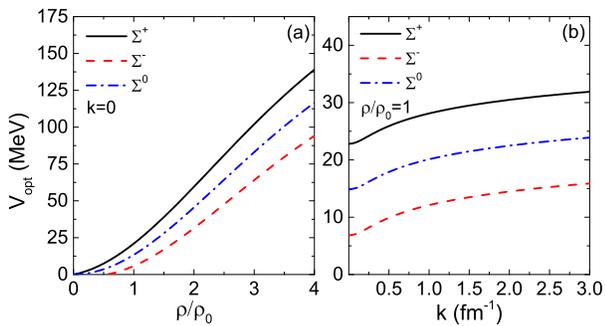}
\caption{ (a) The density and (b) momentum dependence of hyperon-nucleon optical potential. }
\end{figure}

The heavy-ion collisions provide a unique environment for creating the high temperature and dense hadronic matter in laboratories. The strange particles are produced during the compressed stage in two colliding nuclei. The evolutions are deviated by the surrounding baryons via the interaction potential and rescattering process. The dynamics of strange particles is associated with the high-density equation of state, i.e., the rapidity distribution, energy spectra, collective flow etc. Shown in FIG. 2 is the rapidity distribution of $\Sigma$ hyperon in the $^{197}$Au + $^{197}$Au collisions at an incident energy of 2 A GeV. It is obvious that the hyperons are produced mainly in the central rapidity region. At the considered energies, hyperons are mainly created via the nucleon-nucleon and pion-nucleon collisions as the channels $NN \rightarrow NYK$ and $N\pi \rightarrow YK$. The $\Sigma^{+}$ and $\Sigma^{-}$ as the charged hyperon are affected by the Coulomb interaction and optical potential in the nuclear medium. The influence of the optical potential on charged $\Sigma^{\pm}$ is obvious. The hyperons are favorable to escape from the reaction zone instead of being captured by surrounding nucleons with the repulsive potential. The isospin-dependent $\Sigma$N potential influences the threshold energy. Consequently, the lower threshold energy of $\Sigma^{-}$ enhances the yields in nucleon-nucleon collisions. The influence of YN potential and the stiffness of symmetry energy on the final multiplicities of hyperons is listed in TABLE I. The hyperon yields are obviously reduced with the YN potential. The $\Sigma^{-}/\Sigma^{+}$ ratio is nearly independent the potential but weakly modified by the symmetry energy.

\begin{figure}
\includegraphics[width=8 cm]{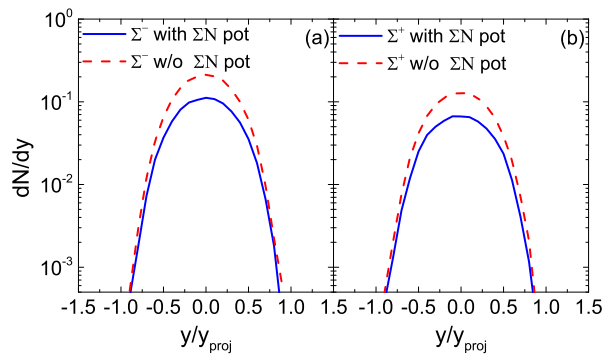}
\caption{ The rapidity distribution of (a) $\Sigma^{-}$ and (b) $\Sigma^{+}$ in the central $^{197}$Au + $^{197}$Au collisions at an incident energy of 2 A GeV. }
\end{figure}

\begin{table*}[!htb]
	\centering
	\caption{The final multiplicity of hyperons and the isospin ratio $\Sigma^{-}/\Sigma^{+}$ in the central $^{197}$Au + $^{197}$Au collisions with different symmetry energy and YN potential.}
	\label{tab1}
	\begin{ruledtabular}
		\begin{tabular}{cccccccccc}
		Beam energy & stiffness of symmetry energy &$\Lambda$ &$\Sigma^{-}$ &$\Sigma^{+}$ &$\Sigma^{0}$ &$\Sigma^{-}/\Sigma^{+}$\\     \hline
	    	1.5 A GeV  &soft (with $\Sigma$N pot.)              &0.1008  &0.0249  &0.0143  &0.0168  &1.736 \\
			1.5 A GeV  &hard (with $\Sigma$N pot.)             &0.1002  &0.0247  &0.0141  &0.0167  &1.754 \\
            2 A GeV    &soft (with $\Sigma$N pot.)               &0.3813  &0.1015  &0.0627  &0.0727  &1.616 \\
            2 A GeV    &hard (with $\Sigma$N pot.)              &0.3777  &0.1011  &0.0619  &0.0726  &1.632 \\   \hline
       Beam energy &YN potential  &$\Lambda$  &$\Sigma^{-}$  &$\Sigma^{+}$  &$\Sigma^{0}$  &$\Sigma^{-}/\Sigma^{+}$\\ \hline
            2 A GeV    &without potential  &0.6294 &0.1694 &0.1052 &0.1232 &1.610 \\
            2 A GeV    &with potential     &0.3652 &0.0918 &0.0571 &0.0683 &1.607 \\
 		\end{tabular}
	\end{ruledtabular}
\end{table*}

The YN potential will influence the phase-space distribution of hyperons produced in the heavy-ion collisions. The kinetic energy spectra of particle emission can be used to extract the in-medium properties of hadrons and the stiffness of symmetry energy. Shown in FIG. 3 is a comparison of the kinetic energy spectra of $\Sigma^{-}$ and $\Sigma^{+}$ with the different case of $\Sigma$N potential and the isospin ratio $\Sigma^{-}/\Sigma^{+}$. It is obvious that the hyperon production is reduced with the potential and the difference of charged hyperons is pronounced. The evolution of charged $\Sigma$ hyperon is influenced by the Coulomb interaction and the optical potential. The elementary cross section for hyperon production is modified via the threshold energy. The reduction of hyperon yields is caused by the enhancement of threshold energy. The in-medium effect exists in the whole energy spectra. Hyperons in the high-energy region are more likely to be ejected from the fire ball formed in heavy-ion collisions and can be probes for investigating the properties of dense baryonic matter. The hunting of sensitive observables for probing the high-density symmetry energy in heavy-ion collisions or from the merging of binary neutron stars was attracted much attention. Particles emitted during the compression stage in heavy-ion collisions are promising observables. In earlier studies, Li applied the pion production in heavy-ion collisions for extracting the equation of state of asymmetric nuclear matter, and proposed the $\pi^{-}/\pi^{+}$ ratio for extracting the high-density behavior of symmetry energy \cite{Li36}. Moreover, the feasibility of studying high-density symmetry energy by $\Sigma^{-}/\Sigma^{+}$ ratios was also investigated by UrQMD transport model \cite{Li37}. Different with the pions, which might be produced in the low and high density region, amount of $\Sigma$ hyperons are mainly created in the high-density domain in nucleus-nucleus collisions. It may be a better observable for probing the high-density symmetry energy. The kinetic energy spectra of the isospin ratio $\Sigma^{-}/\Sigma^{+}$ with different symmetry energy is shown in the down panels. The solid and dashed lines are fitting from the calculations for guiding eyes. It is roughly 5$\%$ difference between the hard and soft symmetry energies at kinetic energies above 700 MeV, which is caused from both the threshold energy correction and transportation of hyperons and nucleons. The $\Sigma^{-}/\Sigma^{+}$ ratio could be a good probe for extracting the behavior of high-density symmetry energy. Further experiments are expected at the HIAF facility in the near future.

\begin{figure}
\includegraphics[width=8 cm]{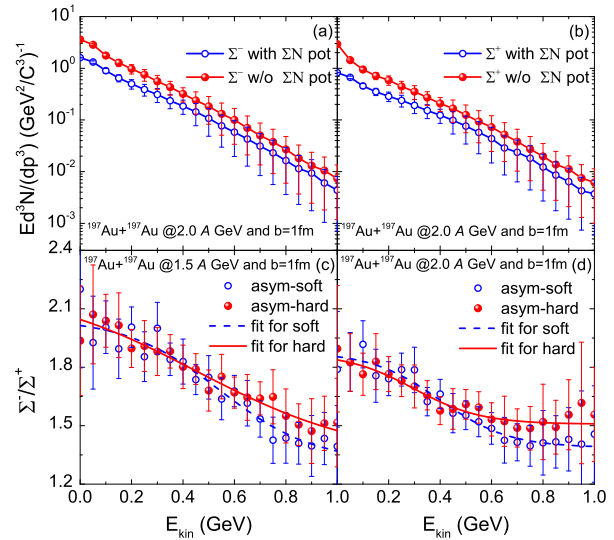}
\caption{\label{fig:wide} The kinetic spectra of (a) $\Sigma^{-}$ and (b) $\Sigma^{+}$ in the reaction of $^{197}$Au + $^{197}$Au at 2 A GeV with the $\Sigma$N potential and the isospin ratio $\Sigma^{-}/\Sigma^{+}$ with the different stiffness of symmetry energy in the reaction of $^{197}$Au + $^{197}$Au at (c) 1.5 A GeV and (d) 2 A GeV, respectively. The lines are shown for guiding eyes.}
\end{figure}

The schematic diagram of directed flow and elliptic flow are given in the Fig. 4. The directed flow shows the degree of deflection of the particles' momentum direction from the beam direction, which is caused from the pressure gradient of nuclear matter formed in heavy-ion collisions. The larger the pressure of compressed nuclear matter is, the larger the degree of deflection of the particles' momentum direction is. Therefore, the directed flow is a sensitive physical quantity which is directly dependent on the nuclear equation of state. The elliptic flow gives the difference between in-plane and out of plane emission of hyperons, and reflects the ability to transform the position space inhomogeneity to momentum space inhomogeneity in heavy-ion collision. The flow information can be expressed as the first and second coefficients from the Fourier expansion of the azimuthal distribution as $\frac{dN}{d\phi}\left(y,p_t\right) = N_0\left[1+2v_1\left(y,p_t\right)cos\left(\phi\right) + 2v_2\left(y,p_t\right)cos\left(2\phi\right)\right]$, where $p_t=\sqrt{p_{x}^{2}+p_{y}^{2}}$ and $y$ are the transverse momentum and the longitudinal rapidity along the beam direction, respectively. The elliptic flow $v_2=<(p^{2}_x-p^{2}_y)/p^{2}_t>$ manifests the competition between the in-plane ($v_2>$0) and out-of-plane ($v_2<$0, squeeze-out) particle emissions. Calculation are performed in the semi-central collisions with the impact parameters of b=4 fm. The E895 data for the $\Lambda$ in-plane flow at AGS energies \cite{Ch38} are well reproduced with the model. The collective flows of $\Sigma$ hyperons are similar to the structure of $\Lambda$, which manifests the well-known "S" shape for the transverse flow and the small $v_{2}$ values. The YN potential and Coulomb interaction affect the hyperon emission in phase space. The precise measurements on the hyperon flows are helpful for investigating the YN, YY potential and high-density symmetry energy.

\begin{figure*}
\includegraphics[width=16 cm]{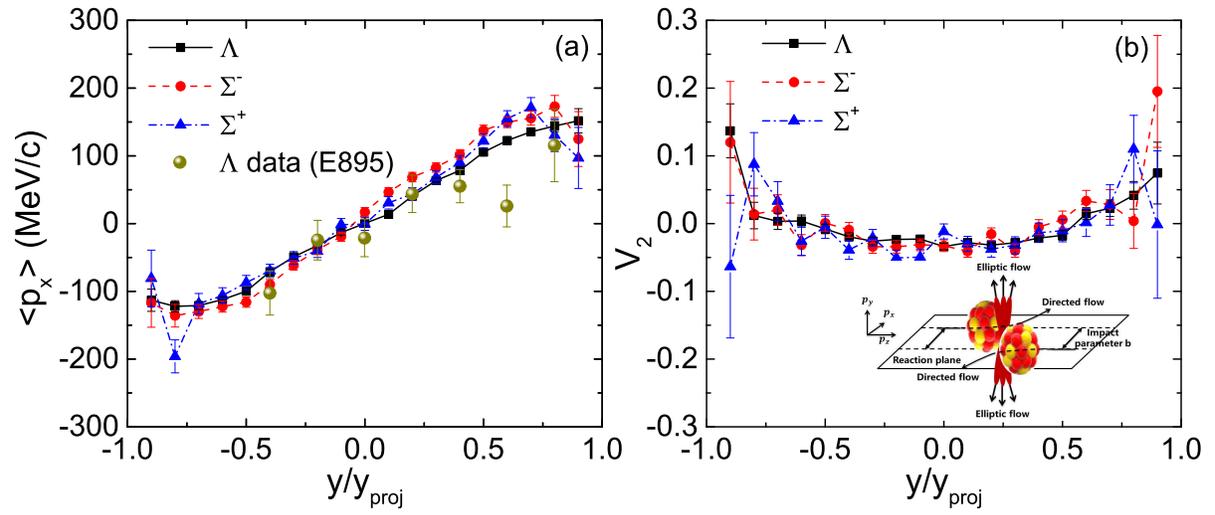}
\caption{\label{fig:wide}  (a) The transverse and (b) elliptic flows of $\Lambda$, $\Sigma^{-}$ and $\Sigma^{+}$ in the $^{197}$Au + $^{197}$Au collisions at 2 A GeV. The experimental data of $\Lambda$ production from E895 Collaboration \cite{Ch38} are shown for comparison. }
\end{figure*}

In summary, the hyperon dynamics and the collective flows in the $^{197}$Au + $^{197}$Au collisions at 2 A GeV have been investigated within the LQMD model. The rapidity distribution and kinetic energy spectra of $\Sigma$ hyperons are calculated. It can be seen that the rapidity distribution and kinetic energy distribution of $\Sigma$ hyperons are sensitive to the hyperon-nucleon potential. The hyperons are produced mainly in the central rapidity region in heavy-ion collisions. The repulsive optical potential increases the threshold energy of hyperons, which results in the decrease of the number of hyperons in middle rapidity region and in the low kinetic energy. The $\Sigma^{-}/\Sigma^{+}$ energy spectra are dependent on the stiffness of symmetry energy. The ratio of $\Sigma^{-}/\Sigma^{+}$ could be a sensitive observable to extract the information of symmetry energy in nuclear matter, especially in high density part. More experiments on the hyperon production in heavy-ion collisions are expected at the HIAF in the near future.Moreover, the collective flows of hyperons are analyzed and, the transverse flow especially, compared with the experimental data from the E895 Collaboration. It is found that the emission of hyperons in the high rapidity region is increased slightly by the repulsive potential.

\end{document}